\newcommand{\btau}{\mbox{\boldmath $\tau $}}
\newcommand{\bPi}{\mbox{\boldmath $\Pi $}}

\newcommand{\bp}{{\bf p}}
\newcommand{\bq}{{\bf q}}
\newcommand{\bk}{{\bf k}}
\newcommand{\ba}{{\bf a}}

\newcommand{\be}{{\bf e}}

\newcommand{\bK}{{\bf K}}
\newcommand{\bA}{{\bf A}}

\newcommand{\atilde}{\tilde{a}}

\newcommand{\Hmath}{\mathcal{H}}

\newcommand{\beq}{\begin{equation}}
\newcommand{\beqn}{\begin{eqnarray}}
\newcommand{\eeq}{\end{equation}}
\newcommand{\eeqn}{\end{eqnarray}}
\newcommand{\nn}{\nonumber}

\documentclass[aps,prl,twocolumn,floatfix,showpacs]{revtex4}
\usepackage{amssymb}

\usepackage{epsfig}
\usepackage{amsmath}

\begin{document}

\title{High-Energy Limit of Massless Dirac Fermions in Multilayer Graphene
using Magneto-Optical Transmission Spectroscopy}
\author{P. \surname{Plochocka}$^{}$}
\email{Paulina.Plochocka@grenoble.cnrs.fr}
\author{C. \surname{Faugeras}$^{}$}
\author{M. \surname{Orlita}$^{}$}
\author{M.L.  \surname{Sadowski}$^{}$}
\author{G. \surname{ Martinez}$^{}$}
\author{M. \surname{Potemski}$^{}$}
\affiliation{$^{}$Grenoble High Magnetic Field Laboratory, CNRS,
38042 Grenoble, France }

\author{M.O. Goerbig }
\author{J.-N. Fuchs}
\affiliation{ Laboratoire de Physique des Solides, CNRS UMR 8502,
Univ. Paris-Sud, F-91405 Orsay cedex, France}

\author{C. \surname{Berger }$^{}$}
\author{W.A. \surname{de Heer }$^{}$}
\affiliation{$^{}$ $^{}$Georgia Institute of Technology, Atlanta,
Georgia, USA }

\date{\today}

\begin{abstract}
We have investigated the absorption spectrum of multi-layer graphene
in high magnetic fields. The low-energy part of the spectrum of
electrons in graphene is well described by the relativistic Dirac
equation with a linear dispersion relation. However, at higher
energies ($>500$~meV) a deviation from the ideal behavior of Dirac
particles is observed. At an energy of 1.25~eV, the deviation from
linearity is $\simeq40$~meV. This result is in good agreement with
the theoretical model, which includes trigonal warping of the Fermi
surface and higher order band corrections. Polarization-resolved
measurements show no observable electron-hole asymmetry.
\end{abstract}


\maketitle

Graphene, a single sheet of graphite, is a two-dimensional system
which exhibits unique electronic properties mostly related to its
peculiar band structure \cite{Novoselv05,Zhang05,Novoselv06,Geim07}.
The remarkable physics exhibited by graphene has its origin in the
conduction and the valence bands which meet at the two inequivalent
($K$ and $K'$) corners of the Brillouin zone. The electrons in the
vicinity of the Fermi energy do not obey Schr\"{o}dinger's equation,
but should instead be described using the quantum-electrodynamic
Dirac equation for relativistic fermions with zero rest mass. The
electrons have a linear dispersion relation whose slope defines a
Fermi velocity $v_F$. In a relativistic analogy, these electrons
behave as massless Dirac fermions moving at an effective speed of
light $v_F$. This system is of great interest from a fundamental
physics point of view and it has even been suggested that graphene
can be used for bench top quantum electrodynamics experiments
\cite{Novoselv05}, for example to test the Klein paradox
\cite{Katsnelson06}. However, in graphene, considering the carriers
as massless fermions remains an approximation and it is both
important and interesting to verify the limits of this
approximation.

Graphene has been extensively investigated using optical
measurements such as Raman scattering~\cite{Ferrari06, Graf07,
Yan07,Faugeras07}, far-infrared absorption (FIR)~\cite{Sadowski06,
Jiang07}, as well as magneto-photoconductivity~\cite{Deacon07}.
Landau level (LL) spectroscopy is a direct and precise tool to test
the linear dispersion relation in the close vicinity of the $K$ and
$K'$ points of the Brillouin zone. In the presence of a magnetic
field $B$, perfect linearity leads to the observed $\sqrt{Bn}$
spacing for the LLs indexed by the integer $n$. In the low-energy
range of the Dirac cone, the linearity of the dispersion relation is
well preserved~\cite{Sadowski06,Jiang07, Deacon07}. However,
graphene is a solid-state system composed of carbon atoms arranged
in a honeycomb lattice and the linear dispersion in these specific
high symmetry points is only a part of a complicated band structure.
This implies that the analogy with neutrinos, massless Dirac
particles, cannot hold everywhere, and we expect a deviation from a
linear dispersion for high energies of the Dirac cone. Although it
is well established that at low energies, electrons in graphene can
be treated as massless Dirac particles, it is crucial to determine
the limits of this approach.

In this Letter, we probe the limits of the massless Dirac fermion
approximation in graphene by extending the previous
studies~\cite{Sadowski06,Sadowski07} to higher magnetic fields,
and, most importantly, to higher energies. Using magneto-optical
transmission spectroscopy, we present a full LL
spectroscopy in magnetic fields up to 32~T, from the far infrared
to the visible range of energy. Transmission measurements
performed in the near visible provide an access to the high-energy
range ($\leq 1.25$~eV) of the Dirac cone. A significant deviation
from the linear dispersion of ideal Dirac fermions is observed.
The experimental data are compared to a theoretical model which
includes higher-order band terms and a good agreement is obtained.
In addition, the asymmetry between electrons and holes has been
probed using polarization-resolved transmission experiments.

\begin{figure}[tbp]
\begin{center}
\includegraphics[width= 0.9\linewidth]{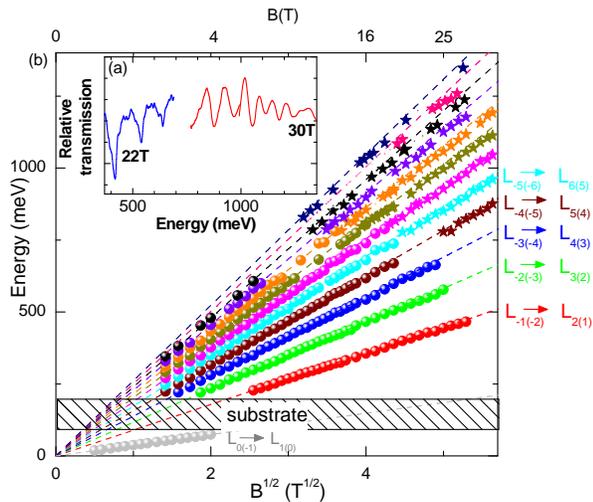}
\end{center}
\caption{(color online) (a) Representative differential
transmission spectra (the $B=0$~T spectra has been subtracted)
measured at the given magnetic fields. (b) Positions of the
absorption lines as a function of the square root of the magnetic
field. Stars represent data obtained in near visible range,
circles denotes the data measured by FTS. Dashed lines are
calculated energy of the transitions between LLs
assuming a linear dispersion. On the right hand side the observed
transition $L_{-m(-n)} \rightarrow L_{n(m)}$ LLs are
denoted. } \label{fig1}
\end{figure}

We have investigated samples containing a high number of graphene
layers (between 70 and 100) grown in vacuum by the thermal
decomposition method, on a (4H) SiC~\cite{Berger04,Berger06}
substrate. Both experiment and theory confirm that the layers are
electronically decoupled so that the system can be considered as a
multi-layer graphene sample
\cite{Hass07,Sadowski06,Sadowski07,Faugeras07}. In particular, these
samples show Raman spectra with the characteristic signature of
single layer graphene \cite{Faugeras07}. It is likely that this
peculiarity of multi-layer graphene, as compared to graphite, is due
to rotational disorder in the stacking, which reduces strongly the
interlayer coupling by roughly two orders of magnitude
\cite{Hass07}. To cover the full spectral range, two different
experiments have been performed. The unpolarized far-infrared
magneto-transmission of the sample at $T=1.9$~K has been measured
using a Fourier Transform Spectroscopy (FTS). To explore the higher
energy range the magneto-transmission up to the visible light range
has been measured at $T=4.2$~K using a tungsten halogen lamp.
Transmission measurements were circular polarization resolved and
spectra recorded for both polarities of the magnetic field. A
representative transmission spectra is shown in Fig.~\ref{fig1}(a).
All spectra show a number of absorption lines which can be assigned
to transitions between $L_{-m(-n)}$ and $L_{n(m)}$ LLs, where $m,n$
enumerates the Landau levels. In the experiment, we observed all
transitions from $L_{-1}\rightarrow L_{2}$ ($L_{-2}\rightarrow
L_{1}$) to L$_{-13}\rightarrow L_{12}$ (L$_{-12}\rightarrow
L_{13}$).

\begin{figure}[tbp]
\begin{center}
\includegraphics[width=0.7\linewidth]{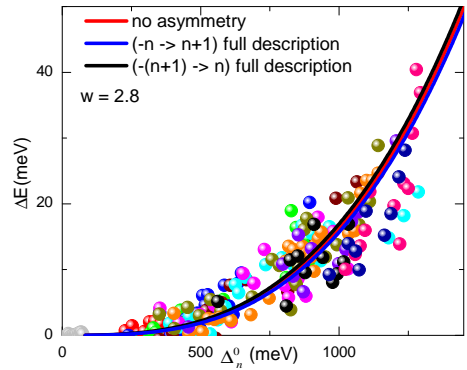}
\end{center}
\caption{(color online) The deviation from linearity $\Delta E$
(from the data in Fig.~\ref{fig1}(b), with the same colors) for
the different transitions as a function of the energy
$\Delta_n^0$. The solid lines is the result of the theoretical
calculation as described in the text.} \label{fig2}
\end{figure}

The energetic position of the observed absorption lines is plotted
as a function of the square root of the magnetic field in
Fig~\ref{fig1}(b). At low energies, the positions of the optical
transitions follow the theoretical (linear) prediction for Dirac
particles [dashed lines in Fig~\ref{fig1}(b)]. For energies above
$\sim500$~meV a deviation from the predicted linear Dirac dispersion
starts to be observed (this can be seen more clearly in
Fig.~\ref{fig2}). Recently Jiang \textit{et al.}~\cite{Jiang07}
presented FIR spectra of a single layer of exfoliated graphene and
determined a Fermi velocity $v_F=1.1\times 10^{6}$m/s. This is
somewhat larger than the value $v_F=1.02\times 10^{6}$m/s obtained
here from the slope in the low-energy region of Fig.~\ref{fig1}(b).
Moreover, in Ref. \cite{Jiang07}, a deviation from the ideal scaling
between adjacent energy transitions was reported, and interpreted as
a consequence of electron-electron interactions. In contrast, in our
data the scaling of the transition energies is well preserved  in
the low energy part of the Dirac cone [see Fig\ref{fig1}(b)], while
at higher energies, we observe deviations due to the non-linearity
of the dispersion relation, as discussed below.

A simple theoretical model has been developed to investigate
deviations from the relativistic Dirac case within the
tight-binding model with a nearest-neighbor (n.n.) hopping term
$t\approx 3$ eV on a honeycomb lattice \cite{Wallace}.
Next-nearest-neighbor (n.n.n.) hopping $t'$, between sites on the
same sublattice, is also included, with $t'/t\sim 0.1$
\cite{Charlier}. The model considers a single graphene layer,
which is a valid assumption in the case of almost decoupled
layers. The energy dispersion for this model may be obtained from
a diagonalization of the $2\times 2$ Hamiltonian matrix, which
reflects the presence of two triangular sublattices $A$ and $B$,
\beq\label{eq03}
\Hmath(\bq)=\left(
\begin{array}{ll}
h'(\bq) & h^*(\bq) \\
h(\bq) & h'(\bq)
            \end{array}
\right), \eeq
with $h(\bq)\equiv
-t\sum_{j=1}^{3}\exp(i\bq\cdot\ba_j)$ and $h'(\bq)\equiv
2t'\sum_{j=1}\cos(\bq\cdot\btau_j)$. Here, the vectors
$\ba_1=\atilde(\sqrt{3}\be_x+\be_y)/2$,
$\ba_2=\atilde(-\sqrt{3}\be_x+\be_y)/2$, and $\ba_3=-\atilde
\be_y$ indicate the coordinates of n.n. carbon
atoms, with a distance $\atilde=0.14$ nm, and
$\btau_1=\sqrt{3}\atilde\be_x$,
$\btau_2=\sqrt{3}\atilde(\be_x+\sqrt{3}\be_y)/2$, and
$\btau_3=\sqrt{3}\atilde(-\be_x+\sqrt{3}\be_y)/2$ those between
n.n.n.

In order to obtain the low-energy spectrum of the dispersion, one
expands $h(\bq)$ and $h'(\bq)$ around the $K$ and $K'$ points at the
edges of the first Brillouin zone (BZ), characterized by the wave
vectors $\pm\bK=\pm(4\pi/3\sqrt{3}\atilde)\be_x$. An expansion in
$\bk=\bq\mp\bK$, up to third order yields \beqn\label{eq04a}
h(+,\bk)&=&\hbar v_F\left(k-\frac{\atilde w_1}{4}k^{*2}-\frac{\atilde^2 w_2^2}{8}|\bk|^2k\right)\\
\label{eq04b} h(-,\bk)&=&-\hbar v_F\left(k^*+\frac{\atilde
w_1}{4}k^2-\frac{\atilde^2 w_2^2}{8}|\bk|^2k^*\right), \eeqn for the
$K$ ($\alpha=+$) and $K'$ ($\alpha=-$) points, respectively, where
we have used the complex notation $k=k_x+ik_y$ and
$v_F=3t\atilde/2\hbar$. Furthermore, we have introduced the
phenomenological parameters $w_1$ and $w_2$, in order to account for
corrections beyond the simplest tight-binding model \cite{trigwarp},
which yields $w_1=w_2=1$. For the n.n.n. term, expanded to lowest
non-trivial order around $K$ and $K'$, one obtains \beq\label{eq05}
h'(\bk)=-3t'+\frac{9t'\atilde^2}{4}(k_x^2+k_y^2) \eeq and thus the
total energy dispersion, taking into account both higher-order band
corrections and n.n.n. hopping, \beqn\label{eq06} \nn
\varepsilon_{\sigma=\pm}^{\alpha=\pm}(\bk) &=&\hbar
v_F\left\{\sigma|\bk|\left[1-\alpha\frac{\atilde
w_1|\bk|}{4}\cos3\phi_{\bk}
\right.\right.\\
\nn
&&-\left.\left.\frac{\atilde^2|\bk|^2}{32}\left(4w_2^2-w_1^2+w_1^2\cos^23\phi_{\bk}
\right)\right]\right.\\
&&\left.+\frac{3t'}{2t}\atilde|\bk|^2\right\}, \eeqn where
$\sigma=+$ denotes the conduction and $\sigma=-$ the valence band,
and $\tan \phi_{\bk}=k_y/k_x$. We have subtracted the unimportant
constant $-3t'$, redefining the zero-energy position. The cosine
terms in Eq. (\ref{eq06}) indicate that the energy dispersion
becomes anisotropic (trigonal warping) \cite{trigwarp}. One clearly
notices from Eq. (\ref{eq06}) that n.n.n. hopping breaks the
particle-hole symmetry but leaves the Fermi velocity unchanged.

In order to account for the magnetic field, we use the Peierls
substitution, which consists of replacing the wave vector $\bk$ by a
momentum operator in the continuum minimally coupled to the vector
potential $\bA$, $\bk\rightarrow\bPi=\bp+e\bA$ \cite{footnote}. The
operator $\bPi$ may be expressed in terms of harmonic-oscillator
ladder operators, with $[a,a^{\dagger}]=1$, and the Peierls
substitution thus reads \beq\label{eq07} k\rightarrow
i\sqrt{2}l_B^{-1}a^{\dagger}\qquad {\rm and} \qquad k^*\rightarrow
-i\sqrt{2}l_B^{-1}a.\eeq Here, $l_B=\sqrt{\hbar/eB}=26/\sqrt{B[T]}$
nm is the magnetic length, which is large in comparison with
$\atilde$, and the above corrections to the linear energy dispersion
are governed, in the presence of a magnetic field, by the small
parameter $\atilde/l_B$. The substitution (\ref{eq07}), together
with Eqs. (\ref{eq04a})-(\ref{eq05}), allows one to calculate the
energies of the relativistic LLs, which, in the absence of the
trigonal-warping terms, read \beq\label{eq11}
\gamma^2\left[n-\frac{4w_2^2-w_1^2}{8}\left(\frac{\atilde}{l_B}\right)^2
n^2\right]=\left(\varepsilon_n-\gamma\frac{3t'\atilde}{\sqrt{2}tl_B}n
\right)^2. \eeq Here, we have defined $\gamma\equiv\sqrt{2}\hbar
v_F/l_B$ and neglected terms due to the order of the operators $a$
and $a^{\dagger}$ when using the substitution (\ref{eq07}). This is
justified in the large-$n$ (semiclassical) limit. In order to
account for trigonal-warping at leading order, we use perturbation
theory, which is justified because $\atilde/l_B\ll 1$. There is no
contribution at first order since $\langle
n|a^{(\dagger)3}|n\rangle=0$ due to the orthogonality of the
eigenstates $\langle n|n'\rangle=\delta_{n,n'}$. At second order,
one obtains $-\gamma^2w_1^2(\atilde/l_B)^2[3n(n+1)+ 2]/8,$ which
needs to be added to the l.h.s. in Eq. (\ref{eq11}). The fact that
trigonal warping is manifest only at order $(\atilde/l_B)^2$ is due
to the magnetic field, which averages to zero the $\cos 3\phi_{\bk}$
term in Eq. (\ref{eq06}) when summing over the angle $\phi_{\bk}$.

One finally obtains, in the large-$n$ limit, where these corrections
become relevant, the energies of the relativistic LLs for both
valleys, $K$ and $K'$, \beq\label{eq12} \varepsilon_{\sigma,
n}=\gamma \frac{3t'}{\sqrt{2}t}\frac{\atilde}{l_B}n
+\sigma\gamma\sqrt{n}\left\{1-\frac{3w^{2}}{8}\left(\frac{\atilde}{l_B}\right)^2
\left[n+\mathcal{O}(n^0)\right]\right\}, \eeq where
$\mathcal{O}(n^0)$ stands for corrections of order unity, and we
have defined $w^2\equiv (w_1^2+2w_2^2)/3$, which may be measured
experimentally. The LL structure in the presence of n.n.n. hopping
has been discussed before in Ref. \cite{Antonio01}. We have checked
the above result within the semiclassical Onsager quantization
scheme, and a comparison with a numerical solution of the Harper
equation on the honeycomb lattice shows excellent agreement even at
small values of $n$ \cite{petra}.

In our experimental study, we compare the deviation $\Delta
E^{\pm}=\Delta_n^0 -\Delta_n^{\pm}$ between the
interband-transition energies
$\Delta_n^0=\gamma(\sqrt{n+1}+\sqrt{n})$ of the ideal case of
Dirac electrons with linear dispersion and the measured
transitions $\Delta_n^+$ (for $-n\rightarrow(n+1)$) and
$\Delta_n^-$ (for $-(n+1)\rightarrow n$), as a function of
$\Delta_n^0$. In Fig.~\ref{fig2}, we compare these deviations to
the theoretical expectations \beq\label{fitted} \Delta E^{\pm}=\mp
\frac{9t'}{2}\left(\frac{\atilde}{l_B}\right)^2+ \frac{3\atilde^2
w^2}{64\hbar^2v_F}\left(\Delta_n^0\right)^3, \eeq obtained from
Eq. (\ref{eq12}) in the large-$n$ limit and $\Delta_n^\pm = \pm
(\varepsilon_{\pm,n+1} -\varepsilon_{\mp,n})$. Note that
$\Delta_n^0\propto \sqrt{Bn}$ when $n\gg 1$. The solid lines in
Fig.~\ref{fig2} show the theoretical result (\ref{fitted}) with a
fitting parameter $w=2.8$ (compared to $w=1$ in the simplest tight
binding model). This somewhat large value indicates that although
the tight-binding model yields the correct functional form and
order of magnitude of band corrections, it underestimates the
strength of these corrections, in particular the effect of the
trigonal warping. In order to account for this enhanced value in a
theoretical model, one would need to include corrections due to
the overlap of the atomic wave-functions on the different lattice
sites, larger distance hopping, and possibly interaction effects.
One may also speculate that, although the graphene layers are only
weakly coupled, the remaining interlayer coupling might play a
role \cite{Antonio02}.
Indeed, trigonal warping in bilayer graphene and graphite is
dominated by \emph{interlayer} hopping, which could be on the same
order of magnitude as the abovementioned dispersion corrections
\cite{maclure,dresselhaus}.

Eq. (\ref{eq12}) shows that the LLs are not electron-hole symmetric
due to n.n.n. hopping. The resulting asymmetry $\mathcal{A}\equiv
|\Delta_n^{-}-\Delta_n^{+}|$ in optical LL transitions is
$\mathcal{A}=3\sqrt{2}\gamma t'\atilde/ t l_B\simeq 0.08 B[T]$ meV,
which is independent of $n$. At fields as high as 30 T, one
therefore expects an electron-hole asymmetry on the order of 2.5
meV, which roughly corresponds to the thickness of the theoretical
curve in Fig.~\ref{fig2}. The effect is thus beyond the resolution
of our experimental data.

\begin{figure}[tbp]
\begin{center}
\includegraphics[width=0.75\linewidth]{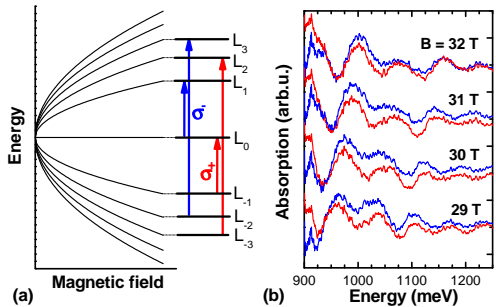}
\end{center}
\caption{(color online) (a) The polarization selection rules for
optical transitions in graphene. (b) The differential transmission
spectra for both circular polarization of the light.} \label{fig3}
\end{figure}

Additional confirmation of the small electron-hole asymmetry in
the LL transitions can be seen using a polarization-resolved
experiment. The polarization selection rules for optical
transition in graphene are shown schematically in
Fig.~\ref{fig3}(a). In Fig.~\ref{fig3}(b) we present transmission
spectra measured for both polarizations at different magnetic
field values. No significant differences between the positions of
the absorption lines can be seen for the different polarization
suggesting that there is no observable asymmetry between the
electron and hole cones.

Symmetry breaking in a gated single sheet of graphene has been
reported recently by Deacon \textit{et al.}~\cite{Deacon07}. In
their cyclotron resonance measurements, the asymmetry is
attributed to n.n. wave-function overlap corrections characterized
by the overlap integral $s_0$ \cite{saito}. In this case the LL
transition asymmetry is $\mathcal{A'}=3\sqrt{2}\gamma
s_0\atilde/l_B\propto B$, which shows that $s_0$ plays a role
similar to $t'/t$, even though the two types of asymmetry have
different microscopic origins. Deacon {\sl et al.} estimated the
strength of the asymmetry to be $\mathcal{A'}\simeq 5$ meV at
$B\simeq 9$~T from the $0\rightarrow1$ and $-1\rightarrow 0$
transitions \cite{Deacon07}, which is in between 5 and 7 times
larger than theoretical estimates, depending on whether one takes
$s_0=0.129$ \cite{saito} or $t'/t \sim 0.1$ \cite{Charlier} with
$t\approx 3$~eV. For a field of 32~T, this would yield an
asymmetry of the order of 18 meV, which should be visible, but is
clearly not observed in the 32T spectra in Fig.~\ref{fig3}(b).


In conclusion, we have probed the high-energy range ($\leq 1.25$~eV)
of the Dirac cone in multi-layer graphene and observed a significant
deviation from the linear dispersion for massless Dirac fermions. A
theoretical model which includes higher order band terms gives good
agreement with experiment. No electron-hole asymmetry in
interband-LL excitations is observed, in agreement with our
theoretical description, where this asymmetry plays a minor role as
compared to trigonal warping of the Fermi surface and higher order
band corrections.

\begin{acknowledgments}
We thank P. Dietl, F. Pi\'echon and G. Montambaux for their
numerical solution of the Harper equation on the honeycomb
lattice. This work was partially supported by contract
ANR-06-NANO-019.
\end{acknowledgments}


\end{document}